\documentclass[aps,prstab,preprint,groupedaddress]{revtex4}
\usepackage{graphics,epsfig}

\begin{document}

\title{Decision Making, Strategy dynamics, and Crowd Formation in
Agent-based models of Competing Populations}

\author{K.P. Chan$^*$, Pak Ming Hui$^*$, and Neil F. Johnson$^+$}
\affiliation{$^*$Department of Physics, The Chinese University of
Hong Kong \\
Shatin, New Territories, Hong Kong \\
$^+$ Department of Physics, University of Oxford, Oxford, OX1 3PU,
UK}

\begin{abstract}
The Minority Game (MG) is a basic multi-agent model representing a
simplified and binary form of the bar attendance model of Arthur.
The model has an informationally efficient phase in which the
agents lack the capability of exploiting any information in the
winning action time series. We illustrate how a theory can be
constructed based on the ranking patterns of the strategies and
the number of agents using a particular rank of strategies as the
game proceeds.  The theory is applied to calculate the
distribution or probability density function in the number of
agents making a particular decision.  From the distribution, the
standard deviation in the number of agents making a particular
choice (e.g., the bar attendance) can be calculated in the
efficient phase as a function of the parameter $m$ specifying the
agent's memory size.  Since situations with tied cumulative
performance of the strategies often occur in the efficient phase
and they are critical in the decision making dynamics, the theory
is constructed to take into account the effects of tied
strategies. The analytic results are found to be in better
agreement with numerical results, when compared with the simplest
forms of the crowd-anticrowd theory in which cases of tied
strategies are ignored.  The theory is also applied to a version
of minority game with a networked population in which connected
agents may share information.

\noindent {\bf Paper to be presented in the 10th Annual Workshop
on Economic Heterogeneous Interacting Agents (WEHIA 2005), 13-15
June 2005, University of Essex, UK.}
\end{abstract}

\maketitle

\section{Introduction}
Agent-based models represent an efficient way in exploring how
individual (microscopic) behaviour may affect the global
(macroscopic) behaviour in a competing population.  This theme of
relating macroscopic to microscopic behaviour has been the focus
of many studies in physical systems, e.g., macroscopic magnetic
properties of a material stem from the local microscopic
interactions of magnetic moments between atoms making up of the
material.   In recent years, physicists have constructed
interesting models for non-traditional systems and established new
branches in physics such as econophysics and sociophysics.  The
Minority Game (MG) proposed by Challet and Zhang
\cite{Challet,Challetbook} and the Binary-Agent-Resource (B-A-R)
model proposed by Johnson and Hui
\cite{BAR1,book,BAR3,preprint,BAR2}, for example, represent a
typical physicists' binary abstraction of the bar attendance
problem proposed by Arthur \cite{Arthur,volatility}. In MG, agents
repeatedly compete to be in a minority group.  The agents have
similar capabilities, but are heterogeneous in that they use
different strategies in making decisions.  Decisions are made
based on the cumulative performance of the strategies that an
agent holds. The performance is a record of the correctness of the
predictions of a strategy on the winning action which, in turn, is
related to the collective behaviour of the agents. Thus, the
agents interact through their decision-making process, creation of
the record of winning actions, and strategy selection process.
Interesting quantities for investigations include the statistics
of the fraction of agents making a particular choice $A(t)$ every
time step and the variance or standard deviation (SD) $\sigma$ of
this number \cite{Challet,book}. These quantities are related in
that knowing the distribution of $A$, one may obtain $\sigma$. The
MG, suitably modified, can be used to model financial markets and
reproduce stylized facts.  The variance, for example, is a
quantity related to the volatility in markets \cite{book}.

Recently, we proposed a theory of agent-based models based on the
consideration of decision-making and strategy dynamics
\cite{NETMG1}. The importance of the strategy selection dynamics
has been pointed out by D'Hulst and Rodgers \cite{rodgers}.  This
approach \cite{rodgers,NETMG1}, which we refer to as the
strategy-ranking theory (SRT), emphasizes on how the strategies
performance ranking pattern changes as the game proceeds and the
number of agents using a strategy in a certain rank for making
decisions.  It is recognized that the SRT has the advantages of
including tied strategies into consideration and avoiding the
troublesome in considering each strategy's performance separately.
The theory, thus, represents a generalization of the
crowd-anticrowd theory \cite{book,preprint,crowd,crowd1} to cases
with tied strategies and strategy ranking evolutions -- two
factors that are particularly important in the so-called
informationally efficient phase of the MG.  The theory has been
applied successfully to explain non-trivial features in the mean
success rate of the agents in (i) MG with a population of
non-networked \cite{rodgers} or networked agents
\cite{NETMG1,NETMG2}, (ii) MG with some randomly participating
agents \cite{RPA}, and (iii) B-A-R model with a tunable resource
level \cite{BAR2}.  In this conference paper, we aim to illustrate
the basic ideas of SRT.  In particular, we present results based
on SRT in evaluating the distribution of $A(t)$ and $\sigma$, in
the efficient phase of MG in non-networked and networked
populations. Validity of the results of our theory is tested
against results obtained by numerical simulations. While the SRT
was developed within the context of MG, many of the ideas are
should also be appliable to a wide range of agent-based models.

\section{Model: The Minority Game}

The basic MG \cite{Challet,Challetbook} comprises of $N$ agents
competing to be in a minority group at each time step. The only
information available to the agents is the history. The history is
a bit-string of length $m$ recording the minority (i.e., winning)
option for the most recent $m$ time steps. There are a total of
$2^{m}$ possible history bit-strings. For example, $m=2$ has
$2^2=4$ possible histories of the winning outcomes: $00$, $01$,
$10$ and $11$. At the beginning of the game, each agent picks $s$
strategies, with repetition allowed. They make their decisions
based on their strategies. A strategy is a look up table with
$2^{m}$ entries giving the predictions for all possible history
bit-strings. Since each entry can either be `0' or `1', the full
strategy pool contains $2^{2^{m}}$ strategies. Adaptation is built
in by allowing the agents to accumulate a merit (virtual) point
for each of her $s$ strategies as the game proceeds, with the
initial merit points set to zero for all strategies. Strategies
that predicted the winning (losing) action at a given time step,
are assigned (deducted) one virtual point. At each turn, the agent
follows the prediction of her best-scoring strategy.  In case of
tied best-scoring strategies, a random choice will be made to
break the tie.

In the present work, we will focus on the regime where $2\cdot
2^{m} \ll N\cdot s$, i.e., the efficient phase.  In MG literature,
a parameter $\alpha = 2^{m}/N$ is defined with $\alpha <
\alpha_{c} \approx 0.34$ characterizing the efficient phase
\cite{Challet1}. Features in this regime is known to be dominated
by the crowd effect \cite{crowd,crowd1}. A quantitative theory in
this regime would have to include the consideration of frequently
occurred tied strategies into account, as the dynamics in this
regime is highly sensitive to the agents' strategy selection.   In
what follows, we introduce the basic physical picture of the
strategy ranking theory and apply it to evaluate the distribution
in the fraction of agents making a particular choice $P(A)$ and
the variance $\sigma^{2}$ from an analytic expression for
non-networked and networked populations.

\section{Numerical and analytical results: Non-networked Agents}

To put our discussions into proper context, we will first present
the numerical results of the quantities that we are focusing on.
Let $A(t)$ be the fraction of agents taking the action ``1" (or
``0") at time step $t$.  As the game proceeds, there will be a
time series $A(t)$.  We may then analyze these values of $A(t)$ by
considering the distribution or probability density function
$P(A)$, where $P(A)dA$ is the probability of having a value within
the interval $A$ to $A+dA$.  In using the MG for market modelling,
$A(t)$ can be taken to be the fraction of agents deciding to buy
(or sell) an asset at time $t$.  In the context of the El Farol
bar attendance problem \cite{Arthur,volatility}, $A(t)$ may be
taken to be the fraction of agents attending the bar. Note that
every realization of the MG may have a different distribution of
strategies among the agents and a different initial bit-string to
start the game. These details do not affect the main results
reported here, especially when we consider cases deep into the
efficient phase, i.e., when $2\cdot 2^{m} \ll N$. To illustrate
the point, we have carried out detailed numerical simulations for
the simplest case of $m=1$ and $s=2$. Figure 1 shows the numerical
results (squares) of $P(A)$ for systems with two different sizes
($N=129$ and $N=4097$), with the aim of emphasizing the size
effect on $P(A)$. Notice that the distribution consists of a few
peaks (five peaks for the case of $m=1$ and $s=2$), indicating
that as the game proceeds the number $A(t)$ jumps among values
characterized by these peak values. For larger population, the
peaks are sharper. Also shown in Fig.1 are the results of the
strategy ranking theory (lines).  The theoretical results are in
reasonably agreement with numerical results.  We defer the
discussion on obtaining the theoretical results to the next
section.

\begin{figure}[tb]
\center{\includegraphics{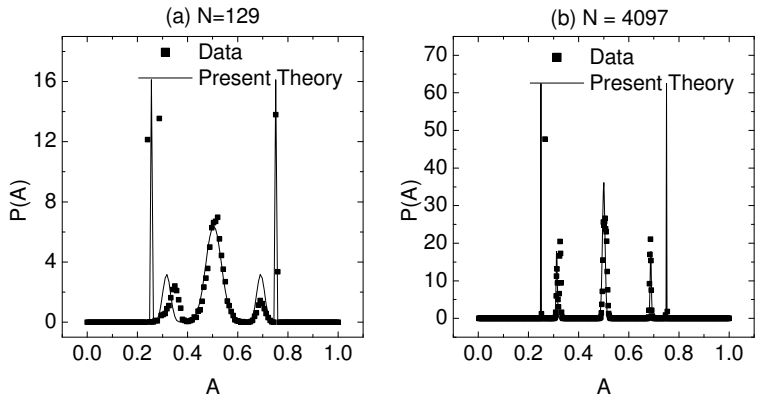}} \caption{The
probability density function $P(A)$ of the fraction of agents
making a particular decision for a typical run with $s=2$ and
$m=1$: (a) $N$=129 and (b) $N$=4097.  The symbols are results
obtained by numerical simulations.  The lines give the results of
the strategy ranking theory.  The peak around $A =0.5$ comes from
the time steps that we have classified as even time steps in the
theory. The peaks on the two sides come from the odd time steps,
where there are registered virtual points in the strategies for
the history bit-string concerned.  For a system of larger size,
the peaks are sharper.} \label{KPChan-fig1}
\end{figure}

Besides the typical results shown in Fig.1, we have studied the
variance $\sigma^{2}$ in the following way.  We carried out
numerical simulations in many realizations using different values
of $m$ and $N$, with $N$ up to $8193$ and $m$ up to $8$.  For each
run, a value of $\sigma^{2}$ is obtained. To facilitate comparison
with theory, we select those data that are deep in the efficient
phase, i.e., with $2\cdot 2^{m}/N < 0.125$ and plotted them (black
dots) in Fig.2 to show the dependence of $\sigma^{2}/N^{2}$ on
$m$. The data points do not show significant scatter, and
essentially fall on a line. Also included in the figure are two
(dashed) lines corresponding to two approximations within the
crowd-anticrowd theory \cite{preprint,crowd,crowd1}.  These
approximations assume that all the strategies can be ranked at
every time step without tied virtual points. One of them assumes
that the popularity rankings, i.e., ranking based on the number of
agents using a strategy, of a strategy and its anti-correlated
partner are uncorrelated and gives an expression for
$\sigma^{2}/N^{2}$ for cases with $s=2$ as \cite{crowd}
\begin{equation}
\frac{\sigma_{flat}^2}{N^{2}} = \frac{1}{24 \times 2^m} \left[1 -
(\frac{1}{2 ^ {m+1}})^2 \right].
\end{equation}
Another approximation is that the ranking of strategies are highly
correlated.  For example, the anti-correlated partner of the
momentarily most-popular strategy is the least-popular one, and so
on. This leads to another expression within the crowd-anticrowd
theory \cite{crowd}:
\begin{equation}
\frac{\sigma_{delta}^2}{N^{2}}  =  \frac{1}{12 \times 2^m} \left[1
- (\frac{1}{2 ^ {m+1}})^2 \right].
\end{equation}
We note that for small values of $m$, the numerical data fall
within the two crowd-anticrowd approximations, with neither of the
approximations capturing the $m$-dependence of
$\sigma^{2}/N^{2}$. As will be discussed later, the strategy
ranking theory gives an {\em analytic} expression for
$\sigma^{2}/N^{2}$ that captures the $m$-dependence very well in
the small $m$ regime where the criteria $2\cdot 2^{m}/N \ll 1$ is
satisfied to a fuller extent.

\begin{figure}[tb]
\center{\includegraphics{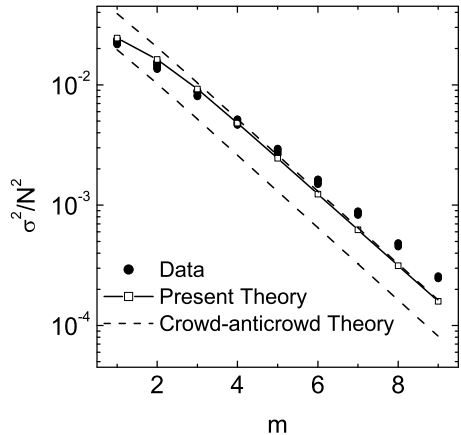}} \caption{The variance
$\sigma^2$/ $N^2$ as a function of $m$, $N$ ranging from $129$ to
$8193$. Only data points satisfying $2\cdot 2^{m}/N < 1/8$ are
shown.  The dashed lines give the two approximations within the
crowd-anticrowd theory.  The open squares give the results, which
are in good agreement with numerical results in the small $m$
regime, of the strategy ranking theory.} \label{KPChan-fig2}
\end{figure}

\section{Strategy ranking theory: Key ideas}
We proceed to discuss how we could obtain the analytic results
shown in Figs.1 and 2, within the strategy ranking theory. Details
of the theory can be found in \cite{rodgers,NETMG1,RPA}. Here we
briefly summarize the key ideas, with the aim to make the theory
physically transparent.  We note that in MG and other agent-based
models of competing populations, it is the interplay between
decision-making, strategy selections, and collective response that
leads to the non-trivial and often interesting global behaviour of
a system.  With this in mind, the strategy performance ranking
pattern is of crucial importance. At any time step, the strategies
can be classified into $\kappa+1$ ranks, according to the virtual
points of the strategies.  The momentarily best-performing
strategy (or strategies) belongs (belong) to rank-1, and so on. At
the beginning of the game, all strategies are tied that thus they
all belong to the same rank. This is also the case when the
strategies are all tied during the game.  Thus, the lower bound of
$\kappa$ is zero.  It is also noted that there are two different
kinds of behaviour in the ranking pattern {\em after} a time step:
(i) the number of different ranks {\em increases} and such a time
step is called an ``even" time step, and (ii) the number of
different ranks {\em decreases} and such a time step is called an
``odd" time step. Take, for example, a time step at which the
strategies are all tied before decision.  Regardless of the
history based on which the agents decide {\em and} the wining
outcome after the agents decided, the strategies split into two
ranks, i.e., $\kappa$ increases from 0 to 1 after the time step.
Half of the strategies belong to the better rank and half to the
worse rank, as half of the strategies would have predicted the
correct winning outcome for the history concerned.  Generally
speaking, the underlying mechanism for this splitting is that {\em
there is no registered virtual point or stored information in the
strategies for the history concerned}.  We call this kind of time
steps ``even" time steps because this is what would happen when
the population encounters a history for decision that had occurred
an even number of times since the beginning of the game, not
counting the one that is currently in use for decisions.

The parameter $\kappa$ has another physical meaning.  It is the
number of history bit-strings that have occurred an odd number of
times since the beginning of the game, regardless the current
history in use for decisions.  Since there are at most $2^{m}$
history bit-strings for a given $m$, the upper bound of $\kappa$
is $2^{m}$.  Thus we have $0 \leq \kappa \leq 2^{m}$.  Therefore,
every time step as the game proceeds can be classified as ``even"
or ``odd", together with a parameter $\kappa$.  For $\kappa=0$
when all the strategies are tied, the time step is necessarily an
even time step.  For $\kappa = 2^{m}$ where there are $2^{m}+1$
ranks, the time step is necessarily an odd time step since all the
histories have occurred an odd number of times, including the
current history in use for decisions.  Noting that the total
number of strategies is $2^{2^{m}}$, there are in general several
strategies in a certain ranking.  In this way, the theory takes
explicit account of cases of tied strategies.

For even time steps (regardless of the value of $\kappa$), there
is no registered virtual points in the strategies for the current
history.  Therefore, {\em even time steps are characterized by
agents making random decisions} \cite{NETMG1,rodgers,NETMG2,RPA}.
Using a random walk argument, the distribution $P_{even,\kappa}(A)
= P_{even}(A)$ is a normal distribution independent of $\kappa$,
with a mean $\mu_{even} = 0.5$ and a variance $\sigma_{even}^{2} =
1/(4N)$, i.e.,
\begin{equation}
P_{even}(A) = \frac{1}{\sqrt{2\pi} \sigma_{even}} \exp\left( -
\frac{(A - \mu_{even})^{2}}{2\sigma_{even}^{2}} \right).
\end{equation}
It turns out that the part of the distribution around $A=0.5$
shown in Fig.1 originates from the even time steps.

For odd time steps, there are registered virtual points or stored
information in the strategies for the current history.  This is
the origin of the crowd effect \cite{preprint,crowd,crowd1}, which
is fundamental to the understanding of collective response in the
class of agent-based models based on MG.  In this case, the
momentarily better performing strategies have predicted the
correct action in the last occurrence of the current history in
use for decision. There will then be more agents using these
better-performing strategies for decisions.  However, the number
is too large, hence forming a crowd, that the winning action in
the last occurrence becomes the losing action in this turn.  This
is the anti-persistent nature or double periodicity of MG
\cite{Challet1,Challet2,Marsili,Savit,Jefferies,Zheng}.  Using the
strategy ranking theory, we know that there are $(\kappa +1)$
ranks among the strategies for time steps labelled $\kappa$. The
ratio of the fractions of strategies in different ranks is given
by \cite{NETMG1} $C_{0}^{\kappa} : C_{1}^{\kappa} : \cdots :
C_{\ell}^{\kappa} : \cdots : C_{\kappa}^{\kappa}$, which are
simply the numbers in the Pascal triangles.  Given that the agents
use their best-performing strategy for decision, we can readily
count the number of agents using a strategy in a particular rank.
As mentioned, the better-performing strategies are more likely to
lead to wrong predictions at odd time steps. This can be modelled
by a winning probability at odd time steps of the form of $(\ell
-1)/\kappa$ for a strategy belonging to rank-$\ell$, for a given
value of $\kappa$ \cite{NETMG1,rodgers}. Putting the information
together, we arrive at the probability density function
$P_{odd,\kappa}(A)$ for $1 \leq \kappa \leq 2^{m}$.  The
distribution $P_{odd,\kappa}(A)$ is given by normal distributions
centered at the mean values of
\begin{equation}
\mu_{odd,\kappa}^{\pm} = 0.5 \pm \frac{C_{\kappa-1}^{2 \kappa
-1}}{ 2 ^ {2 \kappa}}
\end{equation}
with a variance
\begin{equation}
\sigma_{odd,\kappa}^2 = \frac{C_{\kappa-2}^{2 \kappa -2}}{ 2^{2
\kappa - 1}} \frac{1}{4N}.
\end{equation}
Applying Eq.~(4) to the results for $m=1$ in Fig.1, we immediately
identify that the peaks in $P(A)$ at $A=1/4$ and $A=3/4$ are
originated from odd time steps corresponding to $\kappa =1$ and
the peaks at $A=0.6875$ and $0.3125$ are originated from odd time
steps corresponding to $\kappa =2$. These peaks are more
noticeable in Fig.~1(b) when the population size is large.  In
Eq.~(5), the binomial coefficients should formally be expressed in
terms of Gamma functions, so that when the lower index in the
coefficient becomes negative, $\sigma_{odd,\kappa}$ vanishes. This
is the case for $\kappa=1$, and the corresponding distribution
will then be very sharp. This is, for example, the case for the
sharp peaks at $A=1/4$ and $A=3/4$ in Fig.~(1).

To obtain an expression for the overall $P(A)$, including both
even and odd time steps and all possible values of $\kappa$, we
need to take a weighted average over the occurrence of odd and
even time steps \cite{NETMG1}.  The resulting expression is
\begin{equation}
P(A) = \sum_{\kappa=0}^{2^m}
\frac{{C_{\kappa}^{2^{m}}}}{{2^{2^{m}}}}
[(\frac{\kappa}{2^m})P_{odd,\kappa}(A)+(1-\frac{\kappa}{2^m})P_{even}(A)],
\end{equation}
where the factor $C_{\kappa}^{2^{m}}/2^{2^{m}}$ is the probability
of having $\kappa$ history bit-strings occurred an odd number of
times.  The factor $\kappa/2^{m}$ is the probability that given
$\kappa$, the time step is odd.  Applying Eq.~(6) to the case of
$m=1$, we obtain the results (lines) shown in Fig.~1. We note that
the expression in Eq.~(6) is also applicable to $m>1$, as long as
the efficient phase criteria is satisfied.

The calculation of the variance follows from the definition
\begin{equation}
\sigma^{2} = N^{2} \langle (A - \overline{A})^{2} \rangle_{t},
\end{equation}
where $\overline{A}=0.5$ is the mean value of $A$ and the average
$\langle \cdots \rangle_{t}$ represents a time average.  Replacing
the time average by invoking the probability density function
$P(A)$, we have
\begin{eqnarray}
\frac{\sigma^2}{N^{2}} & = & \int_{0}^{1} (A - 0.5)^{2} P(A) dA
\nonumber \\
&=& \sum_{\kappa=0}^{2^m} \frac{{C_{\kappa}^{2^{m}}}}{{2^{2^{m}}}}
\left\{(\frac{\kappa}{2^m})
[\frac{1}{2}(0.5-\mu_{odd,\kappa}^{+})^2 +
\frac{1}{2}(0.5-\mu_{odd,\kappa}^{-})^{2}
+\sigma_{odd,\kappa}^2] + (1-\frac{\kappa}{2^m}) \sigma_{even}^2 \right\}\\
&\approx& \sum_{\kappa=0}^{2^m} \frac{C_{\kappa}^{2^m}}{2^{2^m}}
(\frac{\kappa}{2^m}) (\frac{C_{\kappa-1}^{2 \kappa -
1}}{2^{2\kappa}})^2 \nonumber \\
& = & \sum_{\kappa=0}^{2^m} \frac{C_{\kappa}^{2^m}}{2^{2^m}}
(\frac{\kappa}{2^m}) \left(\frac{1}{2} \prod_{q=1}^{\kappa} (1 -
\frac{1}{2q}) \right)^{2},
\end{eqnarray}
where the approximation is valid for $2.2^m / N << 1$.  Eq.~(9) is
an {\em analytic} expression for $\sigma^{2}$.  The last two
expressions are equivalent and one may use whichever convenient in
obtaining numerical values from Eq.~(9).

Several remarks are worth mentioning.  Firstly, we note that the
expression of $\sigma^{2}$ is closely related to the analytic
expression for the winning probability reported in \cite{NETMG1},
from which an alternative approach arriving at the same result is
possible \cite{thesis}.  Secondly, the results from Eq.~(9) are
plotted (open squares) in Fig.2.  We note that the strategy
ranking theory does capture the $m$-dependence of
$\sigma^{2}/N^{2}$, with good agreement with numerical simulation
results in the range where the criteria $2\cdot 2^{m}/N \ll 1$ is
better fulfilled \cite{rodgers}. The success of the theory stems
from the inclusion of tied strategies, as each rank typically
consists of a number of strategies. In the simplest case of $m=1$,
for example, there are tied strategies in {\em every} time step of
the game. The better agreement with numerical results when
compared with the crowd-anticrowd approximations is thus an
indication of the importance of (i) the tied strategies and (ii)
the time evolution of the ranking pattern from time step to time
step. In MG, both the number of tied strategies, i.e., number of
strategies belonging to the same rank, and the time evolution of
strategy ranking pattern can be readily found.  Thirdly, the
result Eq.~(9) is interesting in that there have been much effort
in trying to re-scale numerical results of $\sigma^{2}$ as a
function of the parameter $\alpha = 2^{m}/N$ so that results from
systems of different values of $N$ and $m$ can be collapsed onto a
single curve.  Eq.~(9) suggests that $\sigma^{2}/N^{2}$ is a
complicated function of $m$, deep in the efficient phase.  In
particular, as one increases the population size at fixed and
small $m$, one should approach the result given by Eq.(9) assuming
a uniform initial distribution of strategies to the agents.  It
is, in fact, possible to include the effects of a finite
population size $N$ into the strategy-ranking theory starting from
Eq.~(8) by incorporating the so-called market impact effects
\cite{rodgers,NETMG2,thesis}.

\section{Networked agents}

\begin{figure}[tb]
\center{\includegraphics{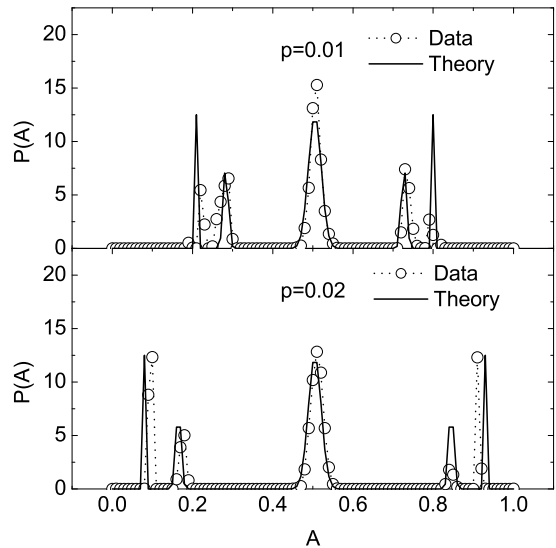}} \caption{ The
probability density function $P(A)$ of the fraction of agents
making a particular decision for a typical run with $s=2$, $m=1$
and $N=1001$ in the networked model of Anghel {\em et al.}
\cite{anghel} for two different values of the connectivity
$p=0.01$ (upper panel) and $p =0.02$ (lower panel). The symbols
are results obtained by numerical simulations.  The lines give the
results of the strategy ranking theory.} \label{KPChan-fig3}
\end{figure}

Systems in the real world are characterized by connected agents
\cite{networks}. The connections are often used for collecting
information from the neighbours.  Recently, several interesting
attempts \cite{NETMG1,NETMG2,anghel,NETMG3,NETMG4} have been made
to incorporate information sharing mechanisms among the agents
into MG and B-A-R models.  As an illustration of the application
of SRT to networked MG, we focus on the model proposed by Anghel
{\em et al.} \cite{NETMG2,anghel}.  As in the MG, Anghel {\em et
al.}'s model \cite{anghel} features $N$ agents who repeatedly
compete to be in a minority group.  Communications between agents
are introduced by assuming that the agents are connected by an
undirected random network, i.e., classical random graph, with a
connectivity $p$ being the probability that a link between two
randomly chosen agents exists.  The links are used as follows.
Each agent compares the cumulated performance of his predictor,
which is the suggested action from his own best-performing
strategy at each time step, with that of his neighbours, and then
follows the suggested action of the best performing predictor
among his neighbours and himself. The $p=0$ limit of the model
reduces to the MG.  Note that the identity of the best-performing
strategy changes over time.  For $p>0$ the predictor's performance
is generally {\em different} from the agent's performance.  It has
been reported that the efficiency of the population as a whole,
characterized by either $\sigma^{2}$ \cite{anghel} or by the
average winning probability per agent per turn \cite{NETMG2},
shows a {\em non-monotonic} dependence on the connectivity $p$
with the most efficient performance occurring at a small but
finite value of $p$.  In other words, a small fraction of links is
beneficial but too many of them are bad.  We have explained the
feature successfully within the framework of SRT \cite{NETMG2}.
The most important point is that, from our understanding of the
non-networked MG (e.g., see Fig.~1), the performance of an agent
actually depends on how similar the $s=2$ strategies that he is
holding, with the best performing ones holding two identical
strategies.  The links then act in two ways depending on the
connectivity.  For low connectivity, the links bring the agents
with two anti-correlated strategies to have the chance to use
other strategies so that these agents will not always join the
crowd at odd time steps and hence with their winning probability
enhanced. For high connectivity, there are so many links that many
agents are linked to the momentarily best-performing predictor or
predictors.  As discussed in previous section, the higher ranking
strategies have a smaller chance of predicting the correct
minority outcome.  When the connectivity is high, there are many
links so that agents have access to strategies that are more
likely to lose.  This leads to a drop in the average winning
probability of the agents \cite{NETMG2}.

Figure 3 shows how the distribution $P(A)$ changes with the
connectivity $p$ at two small values of $p$.  The range of small
$p$ is particularly of interest since for a large population
($N=1001$) the non-monotonic feature occurs for $p < 0.01$. The
symbols (open circles) give the results from numerical
simulations.  The peaks of the distribution $P(A)$ shifts as $p$
is varied.  Applying SRT and incorporating the effects of the
presence of links, we found that $P(A)$ can again be represented
by a weighted sum of distributions characterized by different
kinds of time steps.  In particular, for $p=0.01$, the parameters
of the distributions in Eq.~(4) can be found \cite{NETMG2,thesis}
to be $\mu_{odd,\kappa=1}^{-} = 0.207$, $\mu_{odd,\kappa=2}^{-} =
0.278$, and $\mu_{odd,\kappa=1,2}^{+} = 1 -
\mu_{odd,\kappa=1,2}^{-}$.  The variances are given by Eq.~(5) as
$\sigma^{2}_{odd,\kappa=1}=0$ and $\sigma^{2}_{odd,\kappa=2} =
1/32N$.  Similarly for $p=0.02$, we have $\mu_{odd,\kappa=1}^{-} =
0.075$ and $\mu_{odd,\kappa=2}^{-} = 0.160$, with the same
variances.  The values of these parameters are obtained by
considering the different winning probabilities of the strategies
in different ranks and the change in the number of agents using a
strategy of a certain rank due to the presence of the links.  The
solid lines in Fig.~3 show the distributions obtained by SRT.  The
theory captures the shifts in $P(A)$ with the connectivity $p$.

\section{summary}
In the present work, we illustrated the basic ideas in
constructing a strategy ranking theory for a class of multi-agent
models incorporating the effects of tied strategies and strategy
selections.  We showed how the theory can be applied to MG in the
efficient phase to evaluate the distribution $P(A)$ in the
fraction of agents making a particular decision and the associated
variance $\sigma^{2}$.  In particular, an analytic expression is
given for $\sigma^{2}$ in a non-networked population.  The theory
is also applied to a version of networked MG in which there exists
non-trivial dependence on the performance of the agents as a
function of the connectivity.  Besides $P(A)$ and $\sigma^{2}$,
the theory can also be applied to evaluate other quantities such
as the average winning probability of the agents.  In closing,
while SRT is developed with models based on the MG in mind, the
general approach, namely that of focusing on the ranking pattern
of the strategies and how the pattern evolves in time, should be a
key ingredient in the construction of theories for a large class
of agent-based models.

\acknowledgments{Work at CUHK was supported in part by a Grant
from the Research Grants Council of the Hong Kong SAR Government.
K.P. Chan acknowledges the support of a conference grant from the
Graduate School at CUHK for attending WEHIA 2005.}

\end{document}